\documentclass{emulateapj}

\newcommand{\myemail}{quanz@astro.phys.ethz.ch}

\slugcomment{To appear in ApJ}

\shorttitle{Searching for young Jupiter Analogs around AP Col}
\shortauthors{Quanz et al.}

\begin{document}


\title{Searching for young Jupiter Analogs around AP Col: L-band High-contrast Imaging of the Closest Pre-main-sequence Star}

\author{Sascha P. Quanz$^{1,2}$, Justin R. Crepp$^3$, Markus Janson$^4$, Henning Avenhaus$^2$, Michael R. Meyer$^2$, and Lynne A. Hillenbrand$^3$}
\email{\myemail}


  

\altaffiltext{1}{Based on observations collected at the European Organisation for Astronomical Research in the Southern Hemisphere, Chile, under program number 288.C-5005(A).}
\altaffiltext{2}{Institute for Astronomy, ETH Zurich, Wolfgang-Pauli-Strasse 27, 8093 Zurich, Switzerland}
\altaffiltext{3}{Department of Astrophysics, California Institute of Technology, 1200 E. California Blvd., Pasadena, CA 91125}
\altaffiltext{4}{Department of Astrophysical Sciences, Princeton University, Princeton, NJ, USA}



\begin{abstract}
The nearby M-dwarf AP Col was recently identified by Riedel et al. 2011 as a pre-main-sequence star (age 12 -- 50 Myr) situated only 8.4 pc from the Sun. The combination of its youth, distance, and intrinsically low luminosity make it an ideal target to search for extrasolar planets using direct imaging. We report deep adaptive optics observations of AP Col taken with VLT/NACO and Keck/NIRC2 in the L-band. Using aggressive speckle suppression and background subtraction techniques, we are able to rule out companions with mass $m \geq$ 0.5 -- 1M$_{\rm Jup}$ for projected separations $a>4.5$ AU, and $m \geq 2\,{\rm M_{Jup}}$ for projected separations as small as 3 AU, assuming an age of 40 Myr using the COND theoretical evolutionary models. Using a different set of models the mass limits increase by a factor of $\gtrsim$2. The observations presented here are the deepest mass-sensitivity limits yet achieved within 20 AU on a star with direct imaging. While Doppler radial velocity surveys have shown that Jovian bodies with close-in orbits are rare around M-dwarfs, gravitational microlensing studies predict that  17$^{+6}_{-9}$\% of these stars host massive planets with orbital separations of 1-10 AU. Sensitive high-contrast imaging observations, like those presented here, will help to validate results from complementary detection techniques by determining the frequency of gas giant planets on wide orbits around M-dwarfs.
\end{abstract}




\objectname{AP Col}


\section{INTRODUCTION}


The high-contrast imaging technique has provided some remarkable extrasolar planet discoveries over the past few years \citep[e.g.,][]{chauvin2005,marois2008, marois2010, lagrange2010}. Direct detection allows us not only to infer orbital separations and eventually determine orbits \citep{crepp2011b}, but also to study extrasolar planets in detail through multi-color photometry and spectroscopy \citep{mohanty2007,quanz2010,hinz2010,janson2010,bowler2010}. Investigation of the small sample of directly imaged planets known to date has already revealed unexpected results, including the suggested presence of thick cloud layers \citep{skemer2011}, enhanced metallicity, and indications of non-equilibrium chemistry \citep{barman2011}. Such observations present a challenge to theoretical atmospheric models \citep{skemer2012}.

Dedicated imaging surveys have now searched host stars of all spectral-types, with an initial emphasis on FGK stars \citep{masciardi2005,biller2007,kasper2007,lafreniere2007,chauvin2010,heinze2010}. The fact that the first handful of directly imaged planets have been found orbiting A-stars provides an unambiguous clue that the efficiency of the planet formation process depends upon host star mass. Indeed, extrapolation of the Doppler radial velocity planet population to separations accessible to high-contrast imaging instruments predicts that A-type stars are ideal targets, even though they are intrinsically bright \citep{crepp2011}. Massive stars have a high gas giant planet occurrence rate, and a propensity to form more massive planets with long orbital periods \citep{johnson2007}. While the same line of reasoning suggests a relative paucity of giant planets orbiting M-stars \citep{kennedy2008}, initial results from gravitational microlensing surveys suggest a significantly higher frequency of planets located beyond the snow-line \citep{gould2010,cassan2012} compared to rates of close-in planets inferred from Doppler measurements \citep{johnson2010,bonfils2011}.

High-contrast imaging observations will soon be able to rectify any discrepancies found by indirect planet detection techniques. Planned instruments and new adaptive optics (AO) systems at the VLT \citep{beuzit2006}, Gemini South \citep{macintosh2007}, Palomar \citep{hinkley2011}, the LBT \citep{esposito2012}, and Subaru \citep{tamura2009} will generate unprecedented contrast levels, providing direct access to young planets with mass $m \approx 0.5M_{\rm Jup}$. It is imperative to reach this level of sensitivity because the ``bottom-to-top" formation paradigm of core-accretion \citep{pollack1996} predicts an overall planet occurrence rate that rises steadily with decreasing planet mass. In other words, access to sub-Jovian-mass bodies significantly enhances the prospects for detecting an extrasolar planet around any type of star. 

Youth and distance are important parameters to consider when prioritizing high-contrast imaging targets. With an accurately measured distance of only 8.39$\pm$0.07 pc and estimated age between 12 -- 50 Myr, the nearby M4.5 flare star AP Col (RA: $06^h04^m52^s.16$; DEC: $-34^\circ33'36''.0$ (J2000, E2000)) has recently been identified as the closest pre-main sequence star to the Sun \citep{riedel2011}. Based on its apparent age and space motion, AP Col is most likely a member of the $\sim$40 Myr old Argus/IC 2391 Association \citep{riedel2011}. Stellar companions have been ruled out by lucky imaging observations prior to establishing its proximity and youth \citep{bergfors2010}. Given its (extremely) convenient properties, AP Col is an ideal target to search for gas giant planets by means of direct imaging. 

In this paper, we present the results from two different sets of deep high-contrast imaging observations taken at the VLT and Keck. The observations were carried out in the L-band which, compared to shorter near-infrared (NIR) wavelengths, provides a higher Strehl ratio and more favorable star-planet flux ratio, while also maintaining manageable thermal background levels from the sky and instrument optics. Despite generating the deepest direct imaging mass-sensitivity yet achieved for orbital separations on the scale of the solar-system ($a>3$ AU), we find no evidence for the existence of gas giant planets. We discuss the implications of these observations to the extent that they support the notion that Jovian bodies appear to be rare around M-dwarfs. 


\section{OBSERVATIONS AND DATA REDUCTION}\label{observations_section}
 
\subsection{VLT/NACO}\label{vlt}
Imaging data was acquired on December 4, 2011 UT, with the AO-fed, high spatial resolution camera NACO mounted on ESO's VLT/UT4 8.2-m telescope at Paranal \citep{lenzen2003,rousset2003}. All images were unocculted and taken with the L27 camera (plate scale $\sim$ 27.15 mas pixel$^{-1}$) using the $L'$ filter ($\lambda_{c}=3.8\,\mu$m, $\Delta\lambda=0.62\,\mu$m). We used pupil stabilization mode to enable angular differential imaging (ADI) \citep{marois2006}. 

To enhance the signal-to-noise ratio (S/N) of potential companions, we chose to moderately saturate the stellar PSF core. Unsaturated images were also acquired to calibrate the photometry. Detector reads were recorded individually (``cube mode"). The integration time was set to 0.2s per read (unsaturated reads had an exposure time of 0.1s). We obtained 134 raw data cubes for AP Col in total. Each cube consisted of 64 reads. Between cubes we moved the telescope by $\approx$9$''$ on the sky executing a  5--point dither pattern to facilitate removal of background noise from the sky and instrument optics. Individual detector reads were checked for open loops and poor AO correction. We discarded one complete cube as well as several individual reads from other cubes. The remaining reads were median combined in stacks of 32, yielding 265 science images with an effective integration time of 32$\times$0.2 s each. We measure the PSF FWHM in unsaturated images to be $\sim$4.3 px ($\approx 0.12 ''$), which is comparable to the theoretical diffraction limit, $\Theta\approx\lambda_{c} / D \approx 0.10''$.


Our basic data reduction steps (bad pixel correction, sky subtraction) are described in \citet{quanz2010,quanz2011}. We use the LOCI algorithm to subtract the stellar PSF  \citep{lafreniere2007b}. Following the same naming convention as in the original paper, we set pertinent reduction parameters to the following values: FWHM=4.5 px, $N_\delta$=0.75, $dr$=5, $N_A$=300. 


\subsection{Keck/NIRC2 data}\label{keck}
We also observed AP Col from Keck Observatory on January 7, 2012 UT. Using the 10-m Keck II telescope and AO system \citep{wizinowich2004}, we acquired images with the Near InfraRed Camera (NIRC2; PI: Keith Matthews) in the $L'$ filter
($\lambda_{c} =3.8 \mu m$,  $\Delta\lambda=0.7\,\mu$m). A total of 67 images were obtained, each consisting of 300 reads with 0.106s integration time per read. All images were in the linear detector regime and recorded with the narrow camera setting which provides a plate scale of 9.963 mas pixel$^{-1}$ \citep{ghez2008}. Like the VLT data, we did not use a coronagraph. Vertical angle mode was used to allow for ADI operation. The AO system was running at a refresh rate of 438 Hz. The FWHM  of the PSF was 9.1 px ($\approx 0.09 ''$), i.e., close to the theoretical diffraction limit of $\Theta\approx 0.08''$.

Raw images were processed using standard techniques to remove dark current, flat-field, replace hot-pixel values, and align the target star. In the same manner as described in the previous  section with the VLT data, we used the LOCI algorithm on the 67 images to subtract the stellar PSF. The only difference was the value for the FWHM which was set to 9 px for the Keck data. Table~\ref{observations} shows a summary of relevant observing parameters for both the VLT and the Keck datasets. 

\section{RESULTS}\label{results}
We find no evidence for point sources in either the VLT or Keck data sets, in raw or processed images. Our checks for real companions did initially show potential candidates in the VLT data. However, they subsequently disappeared when combining only a fraction of the final images, and other bright speckles showed up instead. 

\subsection{Derivation of contrast curve}
\begin{figure}
\centering
\epsscale{0.5}
\plotone{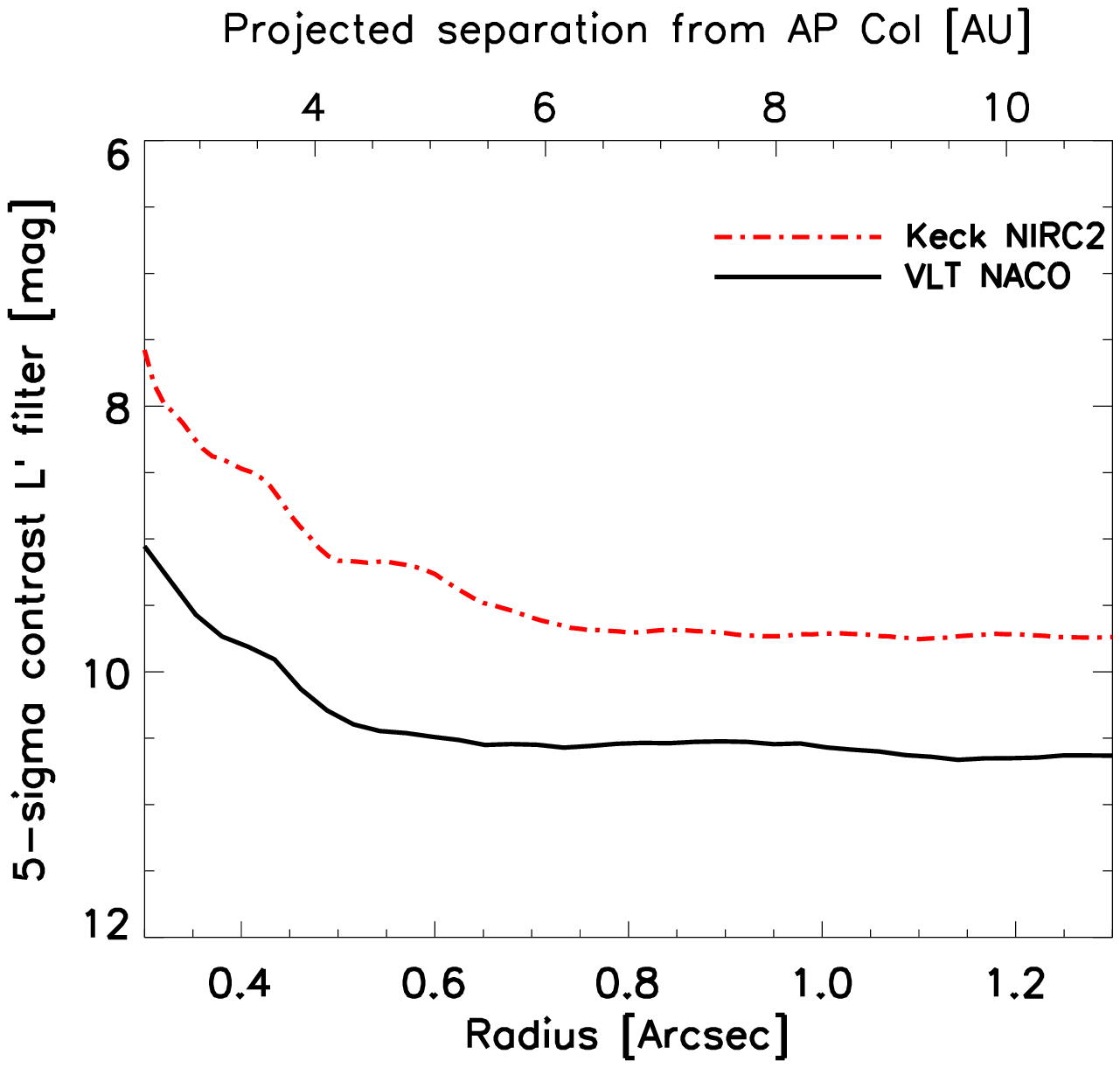}
\caption{Comparison between the 5--$\sigma$ contrast curves obtained with Keck/NIRC2 and VLT/NACO.
\label{comparison}}
\end{figure}

A comparison of the 5-$\sigma$ contrast levels obtained with NACO and NIRC2 is shown Figure~\ref{comparison}. Contrast is defined as:
\begin{equation}
\Delta L'(r) = 2.5 \cdot {\rm log_{10}}\frac{F_*}{5\sigma (r)/\sqrt{\pi r^2_{\rm{ap}}}}
\end{equation}
where $F_{*}$ is the mean stellar flux per pixel in an aperture with radius $r_{\rm{ap}}=3$ pixels (VLT/NACO) or $r_{\rm{ap}}=5$ pixels (Keck/NIRC2), and $\sigma$ is the standard deviation of the pixels within centro-symmetric annuli centered on the star, which are twice as wide as the aperture radius. The VLT/NACO observations provide slightly deeper contrast compared to Keck as a result of lower background levels, more field rotation, and lower airmass. We focus on these data for the remainder of the analysis.

\begin{figure*}
\centering
\epsscale{1}
\plottwo{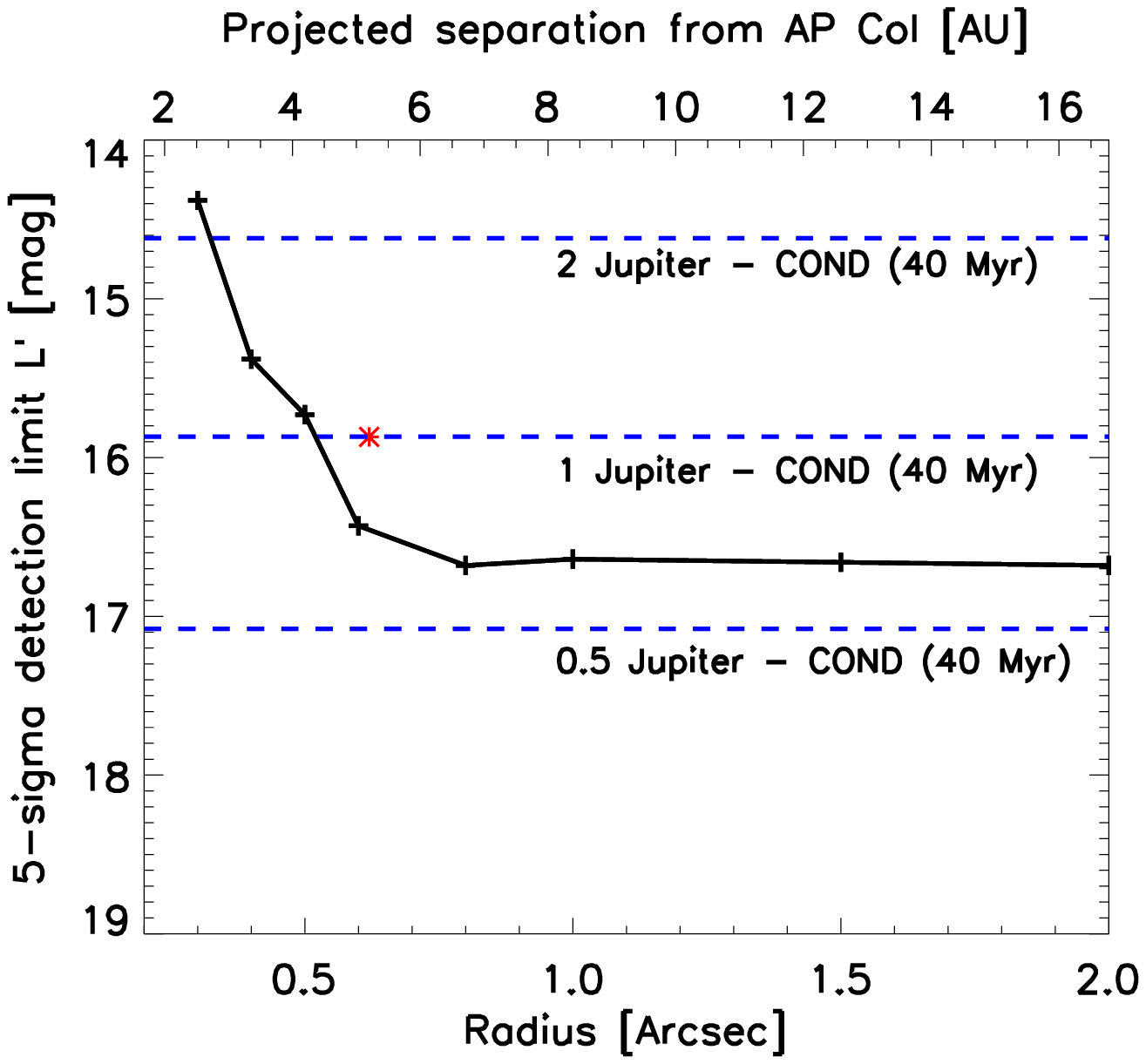}{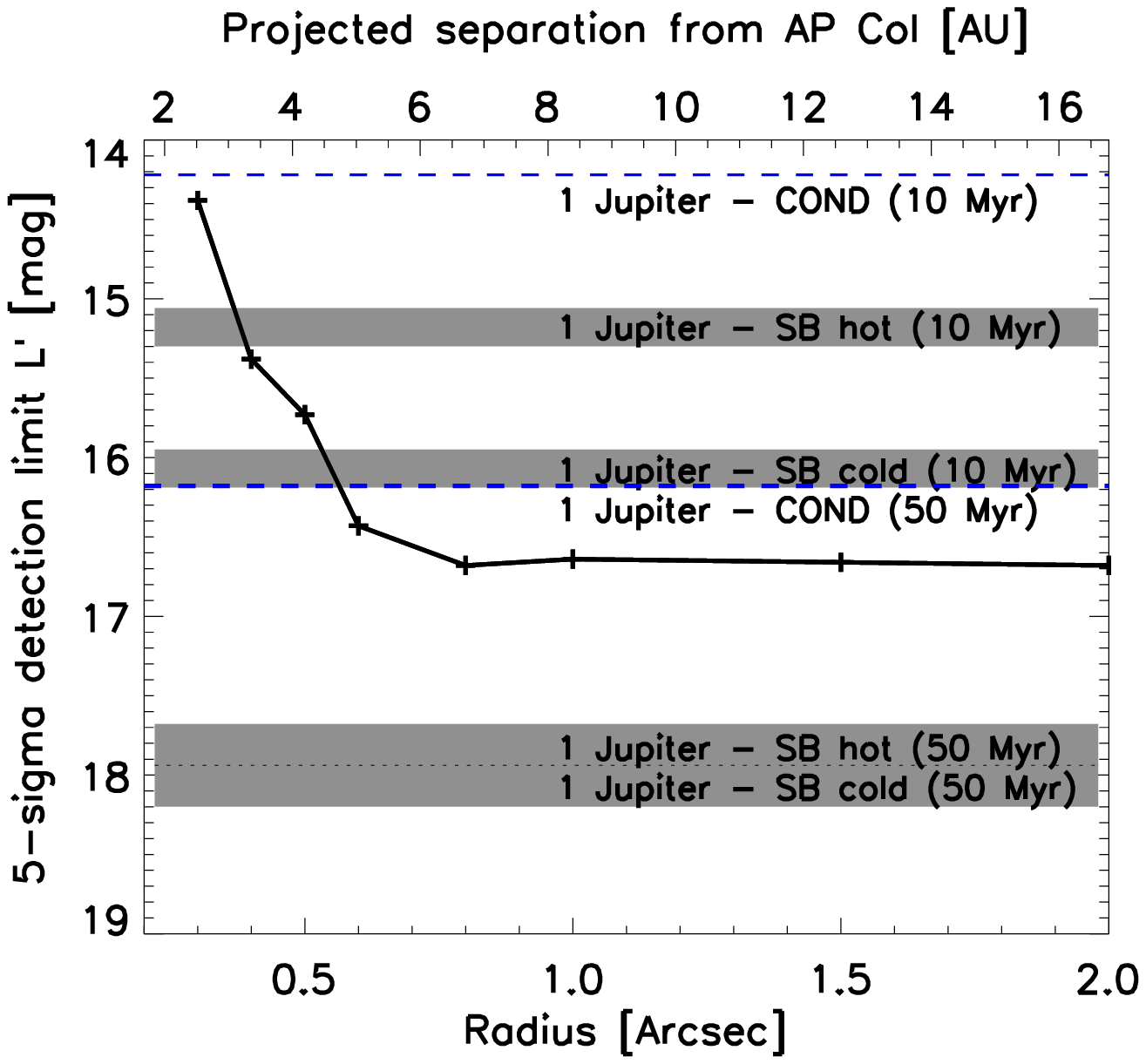}
\caption{5--$\sigma$ detection limit in L$'$ magnitudes for our VLT/NACO data as a function of separation from AP Col (black crosses and curve). \emph{Left:} Overplotted as horizontal dashed lines are predicted L$'$ magnitudes for 40 Myr old objects with masses of 0.5 , 1 and 2 M$_{\rm Jup}$ at the distance of AP Col based on the COND models \citep{baraffe2003}. The red asterisk shows the location of a young Jupiter analog (i.e, 1 ${\rm M_{Jup}}$ at 5.2 AU projected separation).
\emph{Right:} 
Overplotted are predictions from theoretical models for a 1 ${\rm M_{Jup}}$ object with an age of 10 or 50 Myr. The blue, dashed lines are based on the COND models. The grey  shaded areas are predictions from the hot and cold start models from \citet{spiegel2012}, where the spread in magnitude for a given model, i.e., the width of the grey shaded areas,  shows the difference between atmospheres with and without hybrid clouds. All models assume solar metallicity. 
\label{contrast_curve}}
\end{figure*}

To derive our final detection limits, we insert fake planets of known brightness into individual NACO frames. Two planets of the same brightness are inserted at the same radial distance (0.3$''$, 0.4$''$, 0.5$''$, 0.6$''$, 0.8$''$, 1.0$''$, 1.5$''$, and 2.0$''$), but at opposing sides from the central star. The signal-to-noise of the planets in the final LOCI processed image is computed via:
\begin{equation}
\Big(\frac{S}{N}\Big)_{\rm{Planet}}=\frac{F_{\rm{Planet}}}{\sigma/\sqrt{\pi r^2_{\rm{ap}}}},
\end{equation}
where $F_{\rm{Planet}}$ is the measured planet flux in an aperture with radius  $r_{\rm{ap}}=3 $ pixels and  $\sigma$ is the standard deviation of the pixels within an 7 pixel-wide arc between the two fake planets and centered at the same radial distance. By inserting two planets simultaneously and having two arcs for the noise estimate between the planets, we obtain 4 different values for $\big(\frac{S}{N}\big)_{\rm{Planet}}$. To be conservative, we chose the combination yielding the lowest value as our reference. For a given radial separation we lowered the brightness of the planets in steps of 0.5 magnitudes until $\big(\frac{S}{N}\big)_{\rm{Planet}}<5$. The signal-to-noise of the next brightest planet was then used for calibration and the values extrapolated to a level that corresponds to $\big(\frac{S}{N}\big)_{\rm{Planet}}=5$. These values are shown in Figure~\ref{contrast_curve}. 

We emphasize that injecting fake planets of various brightness and retrieving them is the only way to obtain an accurate estimate for the achieved detection limit. LOCI removes flux from potential companions \citep{lafreniere2007b} as a function of separation from the star, LOCI input parameters, and brightness of the companions themselves \citep{pueyo2012}. There are no analytical methods for predicting and correcting for flux losses.


The limiting magnitude in the VLT/NACO data approaches L$'_{\rm limit}\approx16.7$ mag at a separation of $\approx0.7$", beyond which the sensitivity becomes flat indicating that we have reached the floor set by thermal background noise (Fig.~\ref{contrast_curve}). Sensitivity is dominated by subtraction residuals closer to the star where the detection limits are governed more-so by residual scattered light. 

\subsection{Detection limits using COND models}
To convert our on-sky contrast into mass-sensitivity, we must first estimate the $L'$ magnitude of AP Col. Using color transformations provided by the 2MASS team\footnote{http://www.astro.caltech.edu/~jmc/2mass/v3/transformations/}, we estimate AP Col's $K$ magnitude to be $K=6.88\pm0.05$ mag, based on the apparent $K_s$ value of $K_s=6.87\pm0.02$ mag \citep{cutri2003}. A typical M4.5 star has an approximate $K-L'$ color of 0.4 mag \citep{allen2000}. Thus, AP Col has an $L'$ magnitude of $L'\approx6.48\pm0.05$ mag. 

\begin{figure}
\centering
\epsscale{0.5}
\plotone{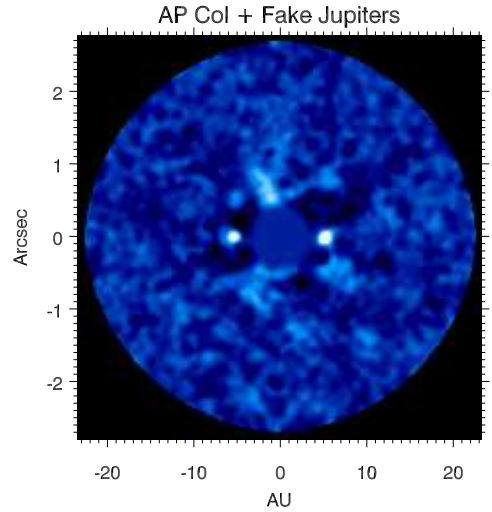}
\caption{Final VLT/NACO image after a fake ``Jupiter" has been inserted in the raw images at two different positions with a projected separation of 5.2 AU from the central star. The brightness of the ``planets" corresponds to a 1 ${\rm M_{Jup}}$ object with an age of 40 Myr based on the COND models. Both ``planets" are clearly detected with with a signal-to-noise of $\sim$9.6 and $\sim$12.1, respectively. The innermost regions that are dominated by subtraction residuals have been masked out.
\label{fake_jupiter}}
\end{figure}

Figure~\ref{contrast_curve} (left-hand panel) compares our detection limits with the predicted L$'$ magnitudes for 40 Myr objects having masses of 0.5, 1 and 2 ${\rm M_{Jup}}$ based on the COND models \citep{baraffe2003}. We select this specific age because \citet{riedel2011} derive a most likely age between 12 -- 50 Myr for AP Col, and membership with the $\sim$40 Myr old Argus/IC 2391 association seems plausible from AP Col's space motion. Our data are sufficiently sensitive to detect objects with masses between 0.5 -- 1 M$_{\rm Jup}$ (or greater) for separations $\ge$4.5 AU.  Inward of 4.5 AU, as close as $\sim$3 AU, objects with masses between 1 -- 2 M$_{\rm Jup}$ (or greater) are detectable. 
Based on these model predictions, \emph{this is the first time that a young Jupiter analog could have been detected by means of direct imaging around another star.} This is also demonstrated in Figure~\ref{fake_jupiter}, where we have inserted two fake planets in the raw images at a projected separation of 5.2 AU and re-run the data reduction. The planets' brightness corresponds to a 1 ${\rm M_{Jup}}$ object with an age of 40 Myr based on the COND models. It clearly shows that we would have detected such an object in our data.

To assess the likelihood of imaging a planet during a single epoch observation, the following relation can be used 
\begin{equation}
P_{\rm detect}(a) = \cos ( \arcsin ( {\rm IWA} / a))\quad,
\end{equation}
where $P_{\rm detect}$ is the detection probability, $a$ the semi-major axis of the planet in AU, and IWA is the inner working angle (in AU) where the contrast curves allows for the detection of this planet. This relation is only valid for circular orbits and assumes uniform sensitivity in azimuth, but takes into account all possible orbital inclinations. For eccentric orbits $P_{\rm detect}$ would be higher for any given $a$, as the planet would spend more time close to apastron where it is easier to detect. 

\begin{figure}
\centering
\epsscale{0.5}
\plotone{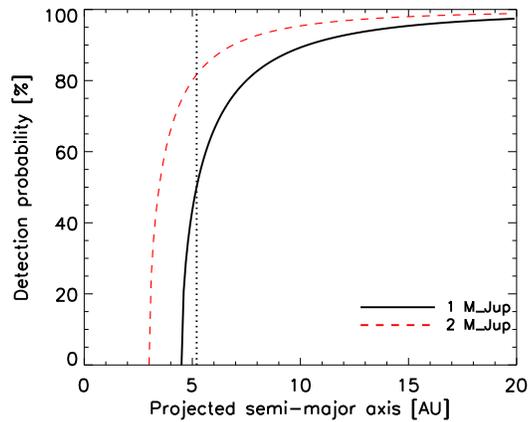}
\caption{Detection probability of a 1 ${\rm M_{Jup}}$ object (black, solid line) and a 2 ${\rm M_{Jup}}$ object (red, dashed line) for our single epoch VLT/NACO observation assuming the detection limits from Figure~\ref{contrast_curve} (left). The vertical, dashed line indicates the position of Jupiter in our Solar System at 5.2 AU.
\label{detect_prob}}
\end{figure}

In Figure~\ref{detect_prob} we plot the detection probability for a 1 ${\rm M_{Jup}}$ planet and a 2 ${\rm M_{Jup}}$ planet using IWA = 4.5 AU for the 1 ${\rm M_{Jup}}$ case and IWA = 3 AU for the 2 ${\rm M_{Jup}}$ case (see, lefthand plot in Figure~\ref{contrast_curve}). It is clear that we had a $\sim$50\% chance to detect (at greater than 5-$\sigma$) a young Jupiter analog at 5.2 AU, assuming one is present. The probability rises quickly for more massive or more distant planets.

\subsection{Comparison of theoretical models}
In the righthand plot of Figure~\ref{contrast_curve}, we have over-plotted in horizontal lines and shaded regions the magnitudes for a 1 ${\rm M_{Jup}}$ object with an age of either 10 or 50 Myr based on theoretical predictions from the COND models \citep{baraffe2003} or more recent models published by \citet{spiegel2012}. The COND models predict that even for an age of 50 Myr we would have been able to detect objects with masses between 0.5 -- 1 M$_{\rm Jup}$ for separations $\ge$5 AU. The models by \citet{spiegel2012}, however, predict that planets are significantly fainter in L$'$ compared to the COND models, regardless of their formation history (hot vs. cold start models). In this case, only for young ages ($\lesssim$20--30 Myr) are our detection limits sufficient to detect a 1 ${\rm M_{Jup}}$ object in the background limit. For an assumed age of 40 Myr only objects with masses $\ge$2 M$_{\rm Jup}$ would have been detected.

\section{DISCUSSION}
Our detection limits are, to our knowledge, the deepest yet achieved for a star targeted by direct imaging in terms of companion mass at a given physical separation. Our results also illustrate an important challenge in interpreting (detections and) non-detections from high-contrast surveys: the systematic and strong dependence upon thermal evolutionary model predictions (see right panel in Figure~\ref{contrast_curve}). The \cite{spiegel2012} models consistently predict fainter L-band magnitudes for an 1 ${\rm M_{Jup}}$ object compared to the \cite{baraffe2003} COND models, most likely as a result of different assumptions regarding atmospheric opacities (Spiegel 2012, private communication). As mentioned in the introduction, presently available atmospheric models fail to self-consistently explain the colors and luminosity of young, low surface gravity objects \citep[e.g.,][]{mohanty2007,skemer2011,barman2011,skemer2012}. 





Nevertheless, our observations demonstrate that AP Col it is unlikely to host a gas giant at separations $\ge$5 AU (Figure~\ref{detect_prob}). For comparison, \citet{delorme2011} have observed 14 nearby, young M-stars, each resulting in non-detections. We note, however, that the highest detection probability of that survey was achieved at separations $\gtrsim$30 AU. Monte Carlo simulations of current and planned high-contrast surveys indeed indicate that low-mass stars are {\it least} likely to harbor a directly detectable planet \citep{crepp2011}. These simulations are based on empirical results from radial velocity studies that find occurrence rates of $\lesssim$2\% for giant planets with close-in orbits \citep{johnson2007, cumming2008, bonfils2011}. 

However, microlensing results suggest that 17$^{+6}_{-9}\%$ of low-mass stars host a gas giant planet with mass between 0.3 -- 10 M$_{\rm Jup}$ at separations between 0.5 and 10 AU \citep{cassan2012}. Thus, the planet frequency seems to rise between the orbital separations probed by Doppler observations and those by microlensing. Combining the currently available results from RV, microlensing, and direct imaging, \citet{quanz2012} have shown that the population of gas giant planets with semi-major axes between $\sim$0.03 -- 30 AU can be described by $d f_{\rm Planet} = M^\alpha a^\beta dM\,da$, with $M$ being the planets' mass, $a$ their semi-major axis, and $\alpha=-1.31$ \citep[cf.][]{cumming2008} and $\beta \lesssim 0.5-0.6$. To further constrain the value for $\beta$ and the orbital separation range where it is valid, additional direct imaging observations effectively probing the regions between $\sim$10 -- 30 AU are required.

\section{Summary \& Conclusions}
Our findings can be summarized as the following:
\begin{itemize}
\item We have used two different telescopes to obtain deep, high-contrast, L-band observations of the nearest (8.4 pc) pre-main sequence star, AP Col, to search for planetary mass companions. Neither dataset revealed the presence of a candidate companion. 

\item Our derived detection limits are the most sensitive yet obtained by direct imaging in terms of planet mass for a given physical separation. For an assumed age of 40 Myr, we could have identified objects with masses $\gtrsim$ 2 M$_{\rm Jup}$ at separations between 3 -- 4.5 AU, and $\gtrsim$ 0.5 -- 1 M$_{\rm Jup}$ for separations $\ge$4.5 AU using the predicted brightness of planets according to the COND models.

\item Accounting for the unknown inclination and position of a potential planet, we had a $\sim$50\% chance to directly image an 1 ${\rm M_{Jup}}$ object at 5.2 AU, provided a young Jupiter analogue exists. 

\item Our detection limits depend on the predicted L-band magnitudes for young, planetary mass objects. Using two different sets of theoretical model predictions, the limits change significantly, demonstrating that: (a) care need be taken for the interpretation of direct imaging results, and (b) empirical model constraints are urgently required.

\end{itemize}

Dedicated imaging surveys for planetary mass companions around M-stars are ongoing \citep[e.g.,][]{delorme2011}. They will eventually provide stringent constraints on the overall frequency and semimajor axis distribution of gas giant planets at separations $\gtrsim$10 AU, complementing the results of RV and microlensing studies. Our results support the notion that future high-contrast imaging programs would maximize their yield by preferentially selecting massive (A, F-type) stars as targets \citep{crepp2011}.

\acknowledgments We thank ESO for granting us Director's Discretionary Time for our VLT/NACO observations and are indebted to Lowell Tacconi-Garman and Julien Girard for their excellent support during the observing run. SPQ thanks D. Spiegel for useful discussions about theroretical models of planetary atmopsheres. 
This research has made use of the SIMBAD database, operated at CDS, Strasbourg, France, and of NASA's Astrophysics Data System.



{\it Facilities:} \facility{VLT:Yepun (NACO)} 
{\it Facilities:} \facility{Keck (NIRC2)}

\clearpage
\hspace{3cm}

\begin{deluxetable}{lll}
\tablecaption{Summary of deep L$'$ imaging observations of AP Col
\label{observations}}           
\tablewidth{\linewidth}
\tablehead{
\colhead{Parameter}  & \colhead{VLT/NACO} & \colhead{Keck/NIRC2} 
}
\startdata
No. of detector reads  $\times$ exp. time  & 64$\times$0.2 s & 300$\times$0.106 s\\
No. of data cubes / stacked images& 134 cubes & 67 images\\
Parallactic angle start / end & -50.11$^{\circ}$ / +59.70$^{\circ}$ & -16.24$^{\circ}$ / +5.59$^{\circ}$\\
Airmass range & 1.02 -- 1.05 & 1.72 -- 1.78 \\
Typical DIMM seeing &  0.7$''$\dots1.0$''$  & $\sim$0.6$''$\\
PSF FWHM\tablenotemark{a} &  $\sim$0.12$''$ (0.10$''$) & $\sim$0.09$''$ (0.08$''$) \\

$\langle{EC}\rangle_{\rm mean}$\tablenotemark{b} & 19.1\% & --- \\ 

\enddata
\tablenotetext{a}{Measured FWHM and theoretical diffraction limited FWHM in parenthesis.}
\tablenotetext{b}{Average value of the coherent energy of the NACO/PSF in data cubes. Calculated by the Real Time Computer of the AO system.}
\end{deluxetable}


\end{document}